\documentclass[11pt]{article}

\date{}

\usepackage{graphicx}

\usepackage{url}
\usepackage{cite}
\usepackage{plain}

\usepackage{wrapfig}
\usepackage{graphics}
\usepackage[english]{babel}
\usepackage[leqno]{amsmath}
\usepackage{amssymb}
\usepackage{verbatim}
\usepackage[mathscr]{eucal}

\usepackage{amstext}
\usepackage{makeidx}
\usepackage{amsthm}

\usepackage{hyperref}
\hypersetup{colorlinks,%
citecolor=red,%
linkcolor=blue,%
urlcolor=black}

\makeindex

\newtheorem{theorem}{Theorem} 

\newtheorem{proposition}{Proprieta'}
\newtheorem{definition}{Definizione}
\newtheorem{notation}{Nota}
\newtheorem{ex}{Esercizio} 
\newtheorem{esempio}{Esempio}

\newcommand{\beq}{\begin{equation}} 
\newcommand{\eeq}{\end{equation}}

\newcommand{\bex}{\begin{ex}} 
\newcommand{\eex}{\end{ex}} 

\newcommand{\bese}{\begin{esempio}} 
\newcommand{\eese}{\end{esempio}} 

\newcommand{\bpro}{\begin{proposition}} 
\newcommand{\epro}{\end{proposition}}

\newcommand{\bthe}{\begin{theorem}} 
\newcommand{\ethe}{\end{theorem}}

\newcommand{\bnote}{\begin{notation}} 
\newcommand{\enote}{\end{notation}}

\newcommand{\bdefi}{\begin{definition}} 
\newcommand{\edefi}{\end{definition}} 

\newcommand{\bc}{\begin{center}} 
\newcommand{\ec}{\end{center}}

\newcommand{\mail}[1]{\href{unina:#1}{\texttt{#1}}}

\usepackage{pdfsync}

\author{ Monica De Angelis \thanks{Univ. of Naples  "Federico II", Dip. Mat. Appl. "R.Caccioppoli", \newline
  80125, Naples, Italy.
\newline\mail{modeange@unina.it},}
}

\title{A note on explicit solutions of  FitzHugh-Rinzel   system   }

\begin{document}
\maketitle


\vspace{3mm}


\begin{abstract}
The numerous scientific feedbacks that the FitzHugh-Rinzel system (FHR) is having in  various scientific fields, lead to further studies
 on   
the determination of its  explicit solutions. Indeed, such a study 
 can help to get a better understanding of  several behaviours in the complex dynamics of biological systems.
In this  note, a class of travelling wave solutions is determined and  specific solutions are achieved to explicitly show  the contribution due to a diffusion term  considered in the FHR model.
\end{abstract}


\section{Introduction}

One of the most commonly known   models in biomathematics is the FitzHugh-Rinzel (FHR) system \cite{k2019,
kl,arxiv}.  It derives from the FitzHugh-Nagumo (FHN) model \cite{i,m2,dr13,krs,drw,dr8,r2,mda13,nono} and unlike the latter, it has an additional variable suitable for evaluating and studying nerve cell bursting phenomena.

In general, bursting oscillations can be described by a system variable that changes periodically from a rapid spike oscillation to a silent phase during which the membrane potential changes slowly \cite{ks}.

Studies concerning bursting phenomena are increasingly present in various scientific fields (see, for instance, {\cite{2020}  and references therein), and in particular,  some applications concern the restoration of synaptic connections. In fact, it seems that certain nanoscale memristor devices
 have the potential to reproduce the behaviour of a biological synapse, suggesting that in the future  electronic synapses may be introduced to directly connect neurons \cite{hc,clag}.

The interest aroused the FHR system applications also leads to the research of explicit solutions. Indeed, in an attempt to understand the various phenomena that the FitzHugh-Rinzel system is able to describe, knowing the expression of the solution can lead to a more complete analysis of the phenomenon itself. In view of this, in this paper, the exact solutions are determined by pointing the research to travelling wave solutions.

The paper is organized as follows. In Section 2,  the mathematical problem is defined. In Section 3, taking into account travelling waves,  a class of explicit solutions is determined, and in Section 4,  a solution has been developed  to  show the incidence of the diffusive term  inserted in the  FHR system. Finally, in Section 5, some concluding remarks have been underlined.

\section{Mathematical Considerations}

  Generally, the FitzHugh-Rinzel   model under  consideration is the  following:

\begin{equation}
\label{11}
  \left \{
   \begin{array}{lll}
    \displaystyle{\frac{\partial \,u }{\partial \,t }} =\, u-u^3/3 + I 
     \,-\, w\,\,+y\,\, \,  \\
\\
\displaystyle{\frac{\partial \,w }{\partial \,t } }\, = \, \varepsilon (-\beta w +c +u)
\\
\\
\displaystyle{\frac{\partial \,y }{\partial \,t } }\, = \,\delta (-u +h -dy),
   \end{array}
  \right.
\end{equation}

\vspace{3mm} \noindent where    $ I, \varepsilon ,\beta,c,d,  h,\delta  $  indicate arbitrary constants.

 \vspace{2mm}In this paper, in order to evaluate also  the contribution due to a diffusion term,  the following  FHR  system  is   considered:

\vspace{3mm}  
\begin{equation}
\label{12}
  \left \{
   \begin{array}{lll}
    \displaystyle{\frac{\partial \,u }{\partial \,t }} =\,  D \,\frac{\partial^2 \,u }{\partial \,x^2 } -au \,\,+k u^2\, (\,a+1\,-u\,)
     \,-\, w\,\,+y\,\,  +I \,  \\
\\
\displaystyle{\frac{\partial \,w }{\partial \,t } }\, = \, \varepsilon (-\beta w +c +u)
\\
\\
\displaystyle{\frac{\partial \,y }{\partial \,t } }\, = \,\delta (-u +h -dy).
   \end{array}
  \right.
\end{equation}

\vspace{3mm}Indeed,   the term with    $ D > 0 $ represents just the diffusion contribution and it derives from the  Hodgkin-Huxley (HH) theory  for nerve membranes when the spatial variation in the potential $ V  $is considered \cite{m2}.

\vspace{2mm} When $ D=0, $ $ k=1/3 $ and $ a=-1, $ system  (\ref{12}) turns into (\ref {11}).

 \vspace{2mm}  After indicating by means of  

\begin{equation}      \label{ic}
u(x,0)\, =\,u_0 \,, \qquad w(x,0)\, =\,w_0  \qquad   y(x,0)= y_0,\qquad\qquad ( \,x\,  \in \Re\,)
\end{equation}

 \vspace{3mm}\noindent   the initial values, from   (\ref{12})   it follows that:

\vspace{3mm}\begin{equation}
\label{14}
\left \{
   \begin{array}{lll}
\displaystyle  w\, =\,w_0 \, e^{\,-\,\varepsilon \beta\,t\,} \,+\, \frac{c}{\beta}\,( 1- e^{- \, \varepsilon \, \beta \,t} )\,+ \varepsilon \int_0^t\, e^{\,-\,\varepsilon \,\beta\,(\,t-\tau\,)}\,u(x,\tau) \, d\tau 
  \\
  \\
\displaystyle  y\, =\,y_0 \, e^{\,-\,\delta\, d\,t\,} \,+\, \frac{h}{d}\,( 1- e^{ - \, \delta \,d  \,t} )\,- \delta \int_0^t\, e^{\,-\,\delta \,d\,(\,t-\tau\,)}\,u(x,\tau) \, d\tau.
 \end{array}
  \right.  
\end{equation}

 \vspace{3mm} So,  when   $ k=1,$ problem (\ref{12})   turns into

	  \begin{equation}                                                     \label{15}
  \left \{
   \begin{array}{lll}
   \displaystyle  u_t - D  u_{xx} + au + \int^t_0 [ \varepsilon e^{- \varepsilon  \beta (t-\tau)}+ \delta e^{- \delta d (t-\tau)} ]u(x,\tau)  d\tau = F(x,t,u) \\    
\\ 
 \displaystyle \,u (x,0)\, = u_0(x)\,  \,\,\,\,\, x\, \in \, \Re, 
   \end{array}
  \right.
 \end{equation} 

}
  \vspace{2mm} \noindent where  

\begin{equation}
\label{188}
\displaystyle F =u^2(a+1-u)+I- w_0 e^{- \varepsilon  \beta  t}+ y_0  e^{- \delta  d  t}- \frac{c}{\beta}( 1- e^{- \varepsilon \beta t} )+\frac{h}{d}( 1- e^{ -\delta d t} ).
\end{equation}
 
  \vspace{2mm} By means of the Laplace transform, the  
 solution of problem  (\ref{15})-(\ref{188})  can be expressed through  an integral equation involving the fundamental solution  $ H(x,t).  $  Indeed, in \cite{2020} it has been proved  that the  
solution assumes the following form:
 
\begin{eqnarray}  \label{A14}
 & \displaystyle \nonumber u (x,t)  =\int_\Re   \,H ( x-\xi, t)\,\, u_0 (\xi)\,\,d\xi \,
 \\
 \\
 &\displaystyle \nonumber\,+\,\int ^t_0     d\tau \int_\Re   H ( x-\xi, t-\tau)\,\, F\,[\,\xi,\tau, u(\xi,\tau\,)\,]\,\, d\xi.
\end{eqnarray}

 \vspace {3mm} Denoting by     $ J_1 (z) \,$  the  Bessel function of first kind and order $\, 1,\,$  and considering  the following  functions:

   \begin{eqnarray} \label {22}
\nonumber \displaystyle  &H_1(x,t)\,=\, \, \, \frac{e^{- \frac{x^2}{4\,D\, t}\,}}{2 \sqrt{\pi  D t } }\,\,\, e^{-\,a\,t}\,+
 \, 
 \\
 \\
&\nonumber \displaystyle - \frac{1}{2} \,\,    \,\,\,\int^t_0  \frac{e^{- \frac{x^2}{4 \, D\, y}\,- a\,y}}{\sqrt{t-y}} \,\, \, \frac{\,\sqrt{\varepsilon} \,\, e^{-\beta \varepsilon \,(\, t \,-\,y\,)}}{\sqrt{\pi \, D \,}} J_1 (\,2 \,\sqrt{\,\varepsilon \,y\,(t-y)\,}\,\,)\,\,\} dy,
   \end{eqnarray}
      
\vspace{5mm}

\begin{equation} \label{H2}
 \displaystyle H_2 =\int _0 ^t  H_1(x,y) \,\,e^{ -\delta d (t-y)} \sqrt{\frac{\delta y}{t-y}}   J_1( \,2 \,\sqrt{\,\delta \,y\,(t-y)\,}\,\,\, dy,
 \end{equation}

 \vspace{3mm} \noindent  one gets

 \begin{equation} \label{A16}
\displaystyle H = H_1  - H_2.
\end{equation}

\section{Explicit Solutions}

Several methods have been developed to find exact solutions of the partial differential equations \cite{rc,dma18,mda19,k18,abg, k2012}. 

Here, in order to find explicit solutions in the form of  travelling solutions,   from system  (\ref{12})
the following equation  is deduced:

\begin{equation} \label{21}
 \displaystyle  u_{tt} = D u_{xxt}  - a u_t+  2 u  u_t (a+1) - 3 u^2 u_t  +\varepsilon \beta w -\varepsilon c -\varepsilon u -\delta u  +\delta h - \delta dy.
\end{equation}

\vspace{2mm} Moreover, letting

\begin{center}
$  \beta \varepsilon = \delta d,  $
\end{center}

 \vspace{2mm} \noindent one obtains

\begin{eqnarray} \label{22}
 \displaystyle  \nonumber &u_{tt} = D u_{xxt}  - a u_t+  2 u  u_t (a+1) - 3 u^2 u_t   -\varepsilon c -\varepsilon u -\delta u  +\delta h  +
\\\
\\
\nonumber   \displaystyle  \nonumber &
\varepsilon \beta (-u_t + D u_{xx}  - a u+ u^2 (a+1) -u^3  +I ). 
\end{eqnarray}

 \vspace{2mm}Now, if one  introduces  the variable wave

\[  z=x- C \,t,\]

\vspace{2mm} \noindent   from   (\ref{22}) one gets

\begin{eqnarray} \label{41}
\nonumber & \displaystyle  D\,C\, u_{zzz} +(C^2-\varepsilon \beta\,D ) u_{zz} - 3 C  u^2 u_{z} +2  C (a+1) u \, u_{z} + \varepsilon \beta u^3 +
\\
\\
\nonumber &  \displaystyle - \, C\, (a\,+ \varepsilon \beta)\, u_z  - \,\varepsilon \beta (a+1) u^2 + \,  \varepsilon \beta \,a \,u  + (\varepsilon +\delta)\,u -K =0.
\end{eqnarray}

 \vspace{2mm} \noindent where 
\begin{center}
$ K = (\delta h-\varepsilon \,c) + \varepsilon \beta \,I. $
\end{center}

\vspace{2mm}The solutions to be determined are of the type

\begin{equation} \label{53}
 u(z) = A\,\, f(z)+\,b \,,
\end{equation}   

\vspace{2mm} \noindent where one assumes  

\begin{equation} \label{A65}
f(z)= \frac{1}{1+e^{(z-z_0)}}.
\end{equation}

\vspace{2mm} Since 
\[ f_{z}- f^2  +f =0, \]

\vspace{2mm} \noindent it results in:
 \begin{eqnarray} \label{55}
 \nonumber & \displaystyle u_z = \,  A \,f ^2(z) - A\,f  
 \\\nonumber
 \\\nonumber
 \nonumber & \displaystyle u_{zz} =  2\,A f^3-3\,A\,f^2  + Af   
 \\\nonumber
 \\\nonumber
  \nonumber & \displaystyle u_{zzz} =  6 A f^4(z) -12  \, A\, f^3(z) +7 A f^2-A f
     \\\nonumber
     \\\nonumber
   & u \, \displaystyle u_z =   A^2\, f^3 + (- A^2  + Ab ) f^2 - Abf   
  \\\nonumber
  \\\nonumber
  \nonumber \displaystyle &u^2  \,u_z  = A^3 f^4+ ( 2 A^2\,b\,- A^3) f^3 +( A b^2 -2 A^2 b ) f^2 -A b^2 f. 
  \end{eqnarray}

\vspace{2mm} In order  to satisfy equation (\ref{41}),  one has to assume

\begin{eqnarray}
 A^2 = 2D 
 \end{eqnarray}

\noindent and  

\begin{equation}
\delta =- \varepsilon.  
\end{equation}

\vspace{2mm} Moreover, under the assumption that 
 
                 \begin{center}$ C>\sqrt{3}/4\, \,  \wedge\,D  >0$

                  \vspace{3mm} or

                \vspace{3mm}  $ \displaystyle C<-\sqrt{3}/4\,\wedge \, D >0 $

   \vspace{3mm} or

\vspace{3mm}  $ \displaystyle 0< D<\, \frac{1}{12}\, ( 3-\sqrt{3} \,\sqrt{3-16\, C^2} \,\,\wedge\, -\sqrt{3}/4\,<\,C \,< \sqrt{3}/4\, $

 \vspace{3mm} or  
 
 \vspace{3mm} $ \displaystyle  D>\, \frac{1}{12}\, ( 3+\sqrt{3} \,\sqrt{3-16\, C^2} \,\, \wedge -\sqrt{3}/4\,<\,C \,< \sqrt{3}/4\,, $

 \end{center}

 \vspace{3mm} \noindent constants $ a,b, $  and $ K  $  must satisfy the following relationships:

 \begin{eqnarray}
&\nonumber\displaystyle b= \frac{1}{6\sqrt{2D}}\,\,\big( \sqrt{2} \,\,\sqrt{2C^2+6D^2-3D}         \,  +2C-6D +3\sqrt{2D} \big)
\\\nonumber\displaystyle
\\\nonumber\displaystyle
 &\nonumber  \displaystyle  a= 3 \, b  - \frac{C}{A} +   \frac{3A}{2} -1
\\\nonumber\displaystyle
\\\nonumber\displaystyle
& \nonumber\displaystyle  K= \varepsilon \beta  b^3  -\varepsilon \beta (a+1) b^2  +[(\varepsilon + \delta)+ \varepsilon \beta a]b.  
\end{eqnarray}

\section{Application}

The previous analysis  allows us  to make some applications. Indeed, in order to point out the contribution of diffusion effects due to  the second order term with  the coefficient  $ D,  $  let us assume, for instance, the following values:

\vspace{3mm}  \[\,C= 1;\,\,\, z_0=0; \,\,\, \varepsilon \,\beta =0.1 \,. \]

\vspace{3mm} In this way, this results in: 

\vspace{3mm}\begin{equation}
b= \frac{\sqrt{6D^2-3D+2}}{6\sqrt{D}}-\frac{\sqrt{D}}{\sqrt{2}}+\frac{1}{3\sqrt{2D}} +\frac{1}{2}
\end{equation}

 \vspace{3mm}\noindent and consequently, one has

\vspace{3mm}\begin{equation}  \label{9}
u(z)= \,    \frac{\sqrt{2D}}{1+e^{z}} \,\,+\,\,\frac{\sqrt{6D^2-3D+2}}{6\sqrt{D}}-\frac{\sqrt{D}}{\sqrt{2}}+\frac{1}{3\sqrt{2D}} +\frac{1}{2}
\end{equation}

 \vspace{5mm} Plotting the graph of function (\ref{9}), it is possible to note how  the diffusion term influences the damping of the solution both when  the coefficient $ D $ is   equal to  or less than $ 1 $ and when  $ D  $ is greater than $ 1. $
 
 \vspace{5mm}
 
 \begin{figure}[h]
\centering
\includegraphics[width=.49\textwidth]{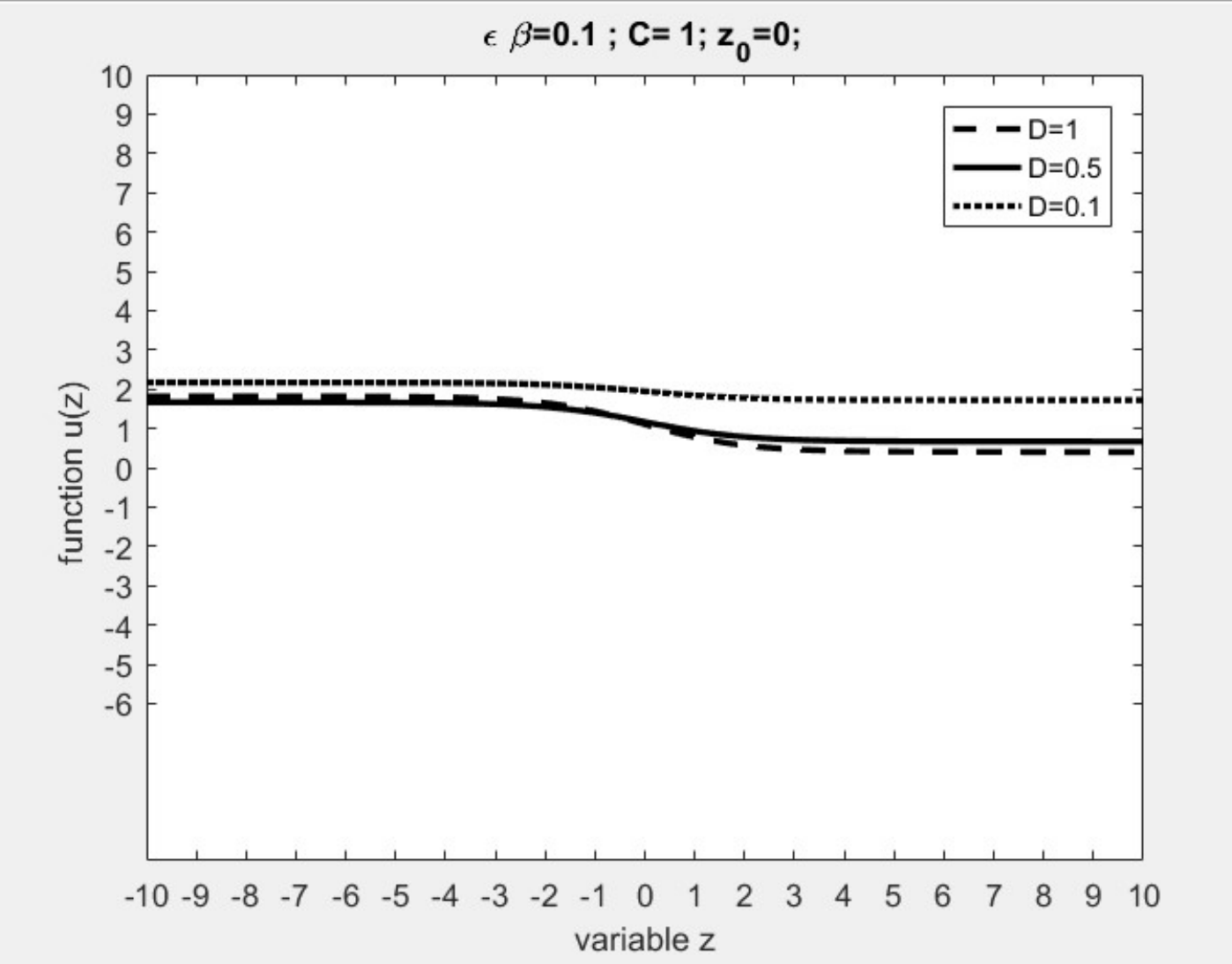} \hfil
\includegraphics[width=.49\textwidth]{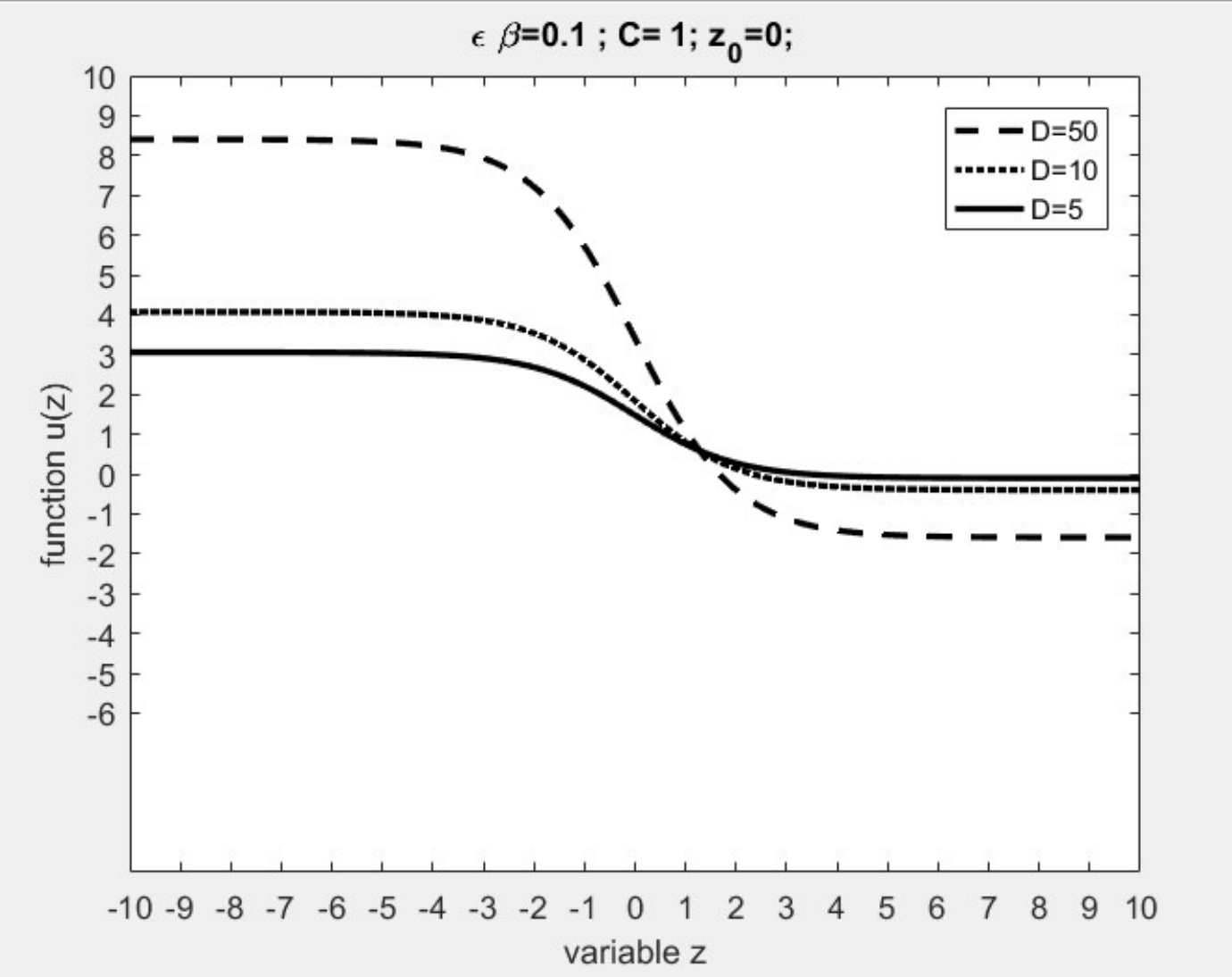}

\caption{Solution u(z) when  $ \varepsilon \beta=0.1, $  $z_0=0  $, $C=1. $  On the left: the values for  the parameter D are such that $ 0<D\leq 1$, while in the right-hand graph: we have considered $ D>1. $ }\label{etichetta}
\end{figure}

\section{Remarks}

\hspace{5mm}$ \bullet $ The paper is concerned with  the ternary autonomous dynamical system of FitzHugh-Rinzel (FHR) which, in biophysics, seems to be  appropriate to describe some phenomena such as bursting oscillations. In this note,  
 the FHR system under consideration includes a diffusion term, represented by a second order term, that derives directly from  the  Hodgkin-Huxley  theory  for nerve membranes and that  is frequently inserted in the  FitzHugh-Nagumo model, too.

\vspace{2mm} $ \bullet $ Solutions can be expressed by means of an integral equation involving the fundamental solution. However, to give  direct feedbacks related to the  contribution due to the  diffusion term $ D $, by means of the method of travelling wave, explicit solutions   have been determined.

\vspace{2mm}  $ \bullet $  Once arbitrary parameters have been set, the trajectories of solutions are shown, whether the parameter $ D $ is less than $ 1 $ or $ D $ is greater than 1.

 \vspace{2mm} $ \bullet $ Of course, as the chosen constants change, the behaviour  of the various solutions can be pointed out.

\section*{Acknowledgment}
The present work has been  developed with the economic support of MIUR (Italian Ministry of University and Research) performing the  activities of the project   
ARS$01_{-}  00861$  “Integrated collaborative 
 systems for smart factory - ICOSAF".

 The  paper has been performed under the auspices of G.N.F.M. of INdAM.


\end{document}